\documentclass[%
 reprint,
 superscriptaddress,
 amsmath,amssymb,
 aps,
floatfix,
]{revtex4-2}
\usepackage{graphicx}
\usepackage{dcolumn}
\usepackage{bm}

\begin{document}

\preprint{APS/123-QED}

\title{Hard X-ray grazing incidence ptychography: Large field-of-view nanostructure imaging with ultra-high surface sensitivity}
\author{P. S. J{\o}rgensen}
\author{L. Besley\footnotemark[\dagger]}

\affiliation{Technical University of Denmark, DTU Energy, 310, Fysikvej, DK-2800 Kgs. Lyngby, Denmark}
\author{A. M. Slyamov}
\affiliation{Xnovo Technology ApS, Galoche Allé 15, 1., Køge 4600, Sjælland, Denmark}
\affiliation{Technical University of Denmark, DTU Energy, 310, Fysikvej, DK-2800 Kgs. Lyngby, Denmark}

\author{A. Diaz}
\affiliation{Paul Scherrer Institut, 111, Forschungsstrasse , 5232 Villigen PSI, Switzerland}
\author{M. Guizar-Sicairos}
\affiliation{Paul Scherrer Institut, 111, Forschungsstrasse , 5232 Villigen PSI, Switzerland}
\affiliation{École Polytechnique Fédérale de Lausanne (EPFL), 1015 Lausanne, Switzerland}

\author{M. Odstr\v{c}il}
\affiliation{Paul Scherrer Institut, 111, Forschungsstrasse , 5232 Villigen PSI, Switzerland}
\affiliation{Carl Zeiss SMT, 22, Carl-Zeiss-Straße, 73447, Oberkochen, Germany}
\author{M. Holler}
\affiliation{Paul Scherrer Institut, 111, Forschungsstrasse , 5232 Villigen PSI, Switzerland}

\author{C. Silvestre}
\author{B. Chang}
\affiliation{Technical University of Denmark, DTU Nanolab, 347, Oersteds Plads, DK-2800 Kgs. Lyngby, Denmark}

\author{C. Detlefs}
\affiliation{European Synchrotron Radiation Facility, 71, avenue des Martyrs, CS 40220,
38043 Grenoble Cedex 9, France}

\author{J. W. Andreasen}
\email{jewa@dtu.dk}
\affiliation{Technical University of Denmark, DTU Energy, 310, Fysikvej, DK-2800 Kgs. Lyngby, Denmark}

\date{\today}

\begin{abstract}

We demonstrate a technique that allows highly surface sensitive imaging of nanostructures on planar surfaces over large areas, providing a new avenue for research in materials science, especially for \textit{in situ} applications. The capabilities of hard X-ray grazing incidence ptychography combine aspects from imaging, reflectometry and grazing incidence small angle scattering in providing large field-of-view images with high resolution transverse to the beam, horizontally and along the surface normal. Thus, it yields data with resolutions approaching electron microscopy, in two dimensions, but over much larger areas and with a poorer resolution in the third spatial dimension, along the beam propagation direction. Similar to grazing incidence small angle X-ray scattering, this technique facilitates the characterization of nanostructures across statistically significant surface areas or volumes within potentially feasible time frames for \textit{in situ} experiments, while also providing spatial information.


\end{abstract}

\maketitle


\section*{Introduction}
\footnotetext{Co-first author with equal contribution}

 
X-ray ptychography is a scanning coherent diffraction imaging technique that retrieves phase and absorption contrast from a series of diffraction patterns collected at various scanning positions with an overlap in sample illumination, which can offer excellent resolution that is not limited by X-ray optics or lenses \cite{Pfeiffer2018,Guizar-Sicairos2021}. Ptychographic data are typically acquired in transmission geometry \cite{Dierolf2010,Holler2014}, however increasing imaging requirements for samples with low-contrast nanoscale features of interest at or near their surface with extended lateral dimensions makes transmission-based imaging challenging \cite{DeCaro2016}. Coherent diffraction imaging of X-rays near grazing incidence has been applied in reflection geometry for the reconstruction of non-periodic surface structures \cite{Sun2012}, however the application of ptychography at grazing incidence with hard X-rays would allow for non-isolated objects of arbitrary size to be imaged in extended samples. As such, ptychography in a reflection geometry is gaining interest and has been explored in the range of EUV wavelengths \cite{Tanksalvala2021,Zhang2015}. 

Ptychographic imaging in reflection geometry has been shown to provide quantitative imaging of nanostructures, sensitive to both chemical and structural contrast, however to date, reflection-mode ptychographic imaging with hard X-rays remains unexplored, with the exception of Bragg-condition ptychography  \cite{Godard2011}, and in crystal truncation rod measurements \cite{Zhu2015}. The application of X-rays in grazing incidence in such an imaging technique would prove an invaluable imaging tool for surface features on the tens of nm scale. This is highly relevant for a variety of technologies with nanoscale surface features where imaging over several hundreds of $\mu$m is required. Critical angles for total reflection of X-rays from surfaces of common materials are typically somewhat less than $\theta_c = 1^\circ$, which results in the beam footprint being elongated by a factor of up to several hundred times the transverse length. The geometry of grazing incidence also presents significant experimental challenges by requiring a highly precise alignment of the sample plane with the scanning plane. Transmission X-ray ptychography requires very precise knowledge of scan positions in order for successful reconstructions, and when grazing incidence geometry is introduced, errors can be amplified by a factor of $1/\sin(\theta)$, placing further requirements on sample alignment and motor precision. Given that critical angles in the hard X-ray regime are typically  ($<1^\circ$), the geometry of grazing incidence presents a significant challenge even when compared to EUV ptychography where incidence angles can be approximately 1 to 2 orders of magnitude larger \cite{Tanksalvala2021}.

We present an extension of the ptychographic X-ray imaging technique by performing it at grazing incidence angles that are typical of the hard X-ray regime and, for the first time, show a successful reconstruction of phase contrast images. The proposed method combines the high-resolution and robustness of ptychographic imaging with the macroscopic probing and flexibility of the grazing incidence geometry, enabling the multi-scale imaging of the morphology of thin-films and surfaces. In the direction parallel to the X-ray beam (longitudinal), the spatial resolution is sacrificed for an increased area of sample, resulting in an image with highly anisotropic resolution. We show that such experiments can be implemented at beam lines used for standard transmission ptychography with the addition of further surface alignment procedures. 
We also demonstrate the simulation of diffraction data from model structures using the multislice technique outlined in \citet{Li2017} which has been shown to accurately model complex multiple-scattering phenomena. Such realistically simulated data is used to validate the accuracy of the reconstruction, which does not use a multislice approach.


\begin{figure*}
  \includegraphics[width=\textwidth]{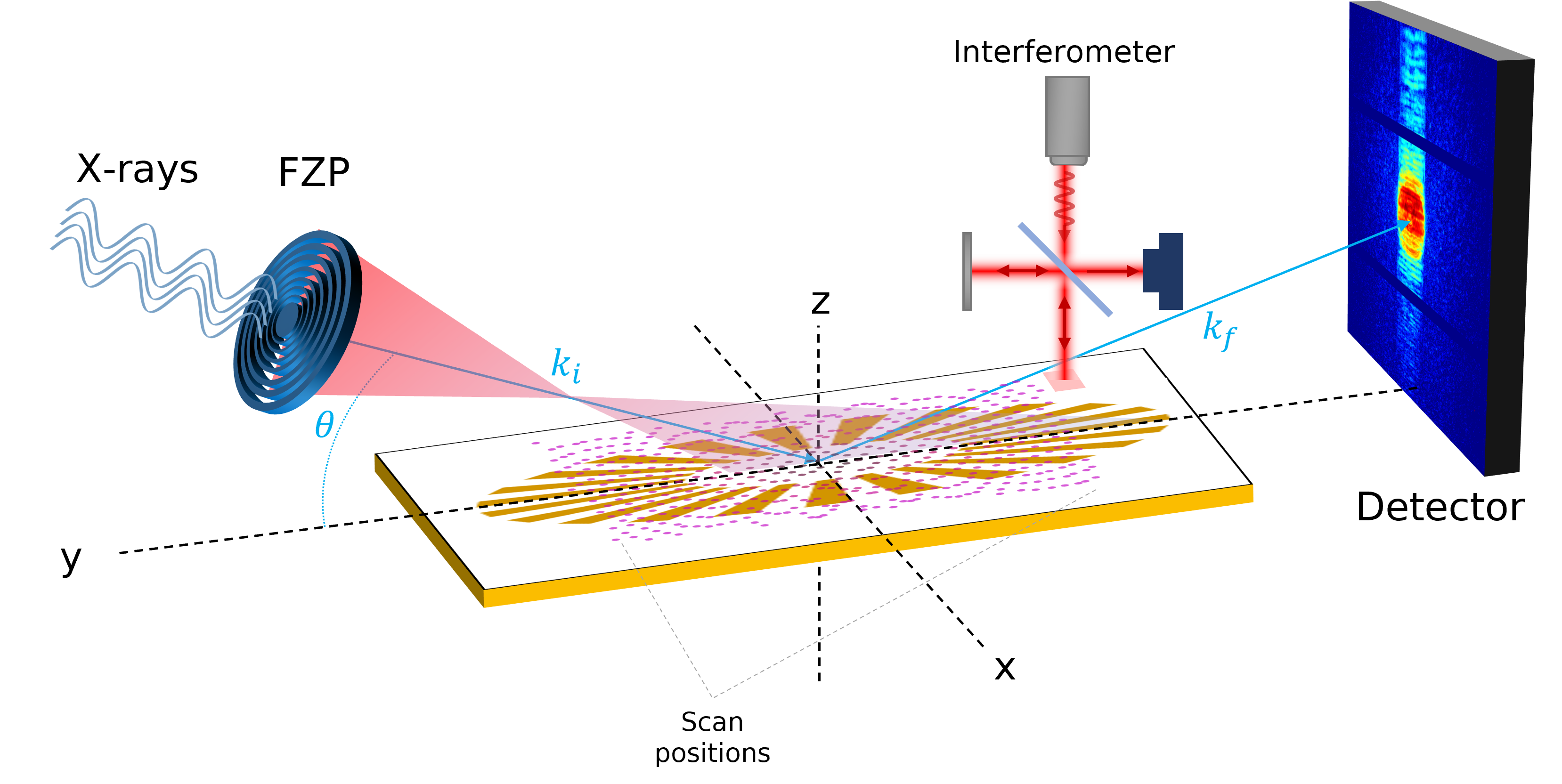}
  \caption{\label{fig:setup} Schematic representation of the experimental set-up. The incident X-rays are initially focused by a Fresnel zone plate (FZP), and the sample is downstream of the focal point. An interferometer is used for aligning the sample surface with the scanning plane. The sample is scanned along its surface with respect to the beam at the positions drawn in pink, in such a way that neighbouring illuminated areas largely overlap, while keeping $z$ constant.}
\end{figure*}

\section*{Results}
\subsection*{Experimental}
\begin{figure}[h]
    \includegraphics[width=\columnwidth]{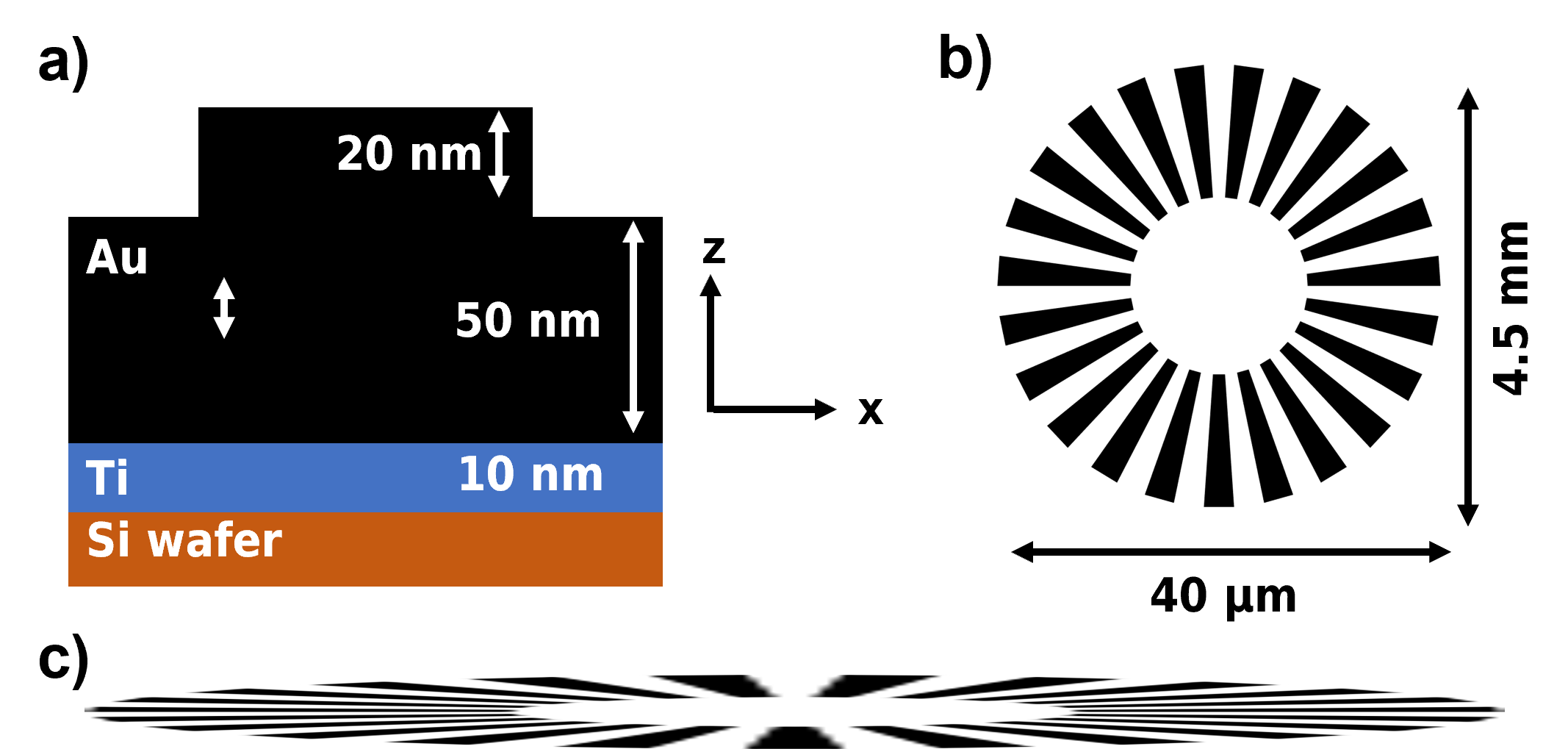}
    \caption{\label{fig:sample} a) a cross section of the test sample orthogonal to the sample plane. A Si wafer is deposited with layers of Ti and Au, forming the substrate. On top of the Au layer, the structure pattern, also Au, is deposited as a 20~nm layer, creating a "Au on Au" structure. b)  Test structure for grazing incidence ptychographic X-ray imaging, an elongated Siemens star. c) The Siemens star test structure from b) but drawn in an aspect ratio closer to the real sample used in experiments }
\end{figure}

We provide a comparison of reconstructed images of samples with their nominal structural design used in the lithographic process, with heights measured by atomic force microscopy (AFM). Fig.~\ref{fig:setup} shows the experimental setup during scanning. Due to the experimental geometry, the reconstructed pixel has a high aspect ratio and the reconstructed images appear squeezed along the grazing-incidence direction. To compensate for this effect and simplify the interpretation of the results, an elongated structure was fabricated for the proof-of-concept experiment. Fig.~\ref{fig:sample}a shows the design of the sample in cross-section. The Si wafer is first coated with a thin titanium adhesion layer. On top of this, a 50~nm layer of Au is deposited, followed by the test structure. Further details of sample fabrication are given in the methods section. The thickness of all structures above the substrate is 20~nm. The stretched "Siemens Star" with truncated spokes (Fig.~\ref{fig:sample}b and \ref{fig:sample}c) has dimensions of 0.04 mm$\times$ 4.5 mm. 

Image acquisition is performed at different angles of incidence $\theta$, in particular above and below the critical angle ($\theta_{c}$) for total external reflection. 

\begin{figure}[h]
    \includegraphics[width=\columnwidth]{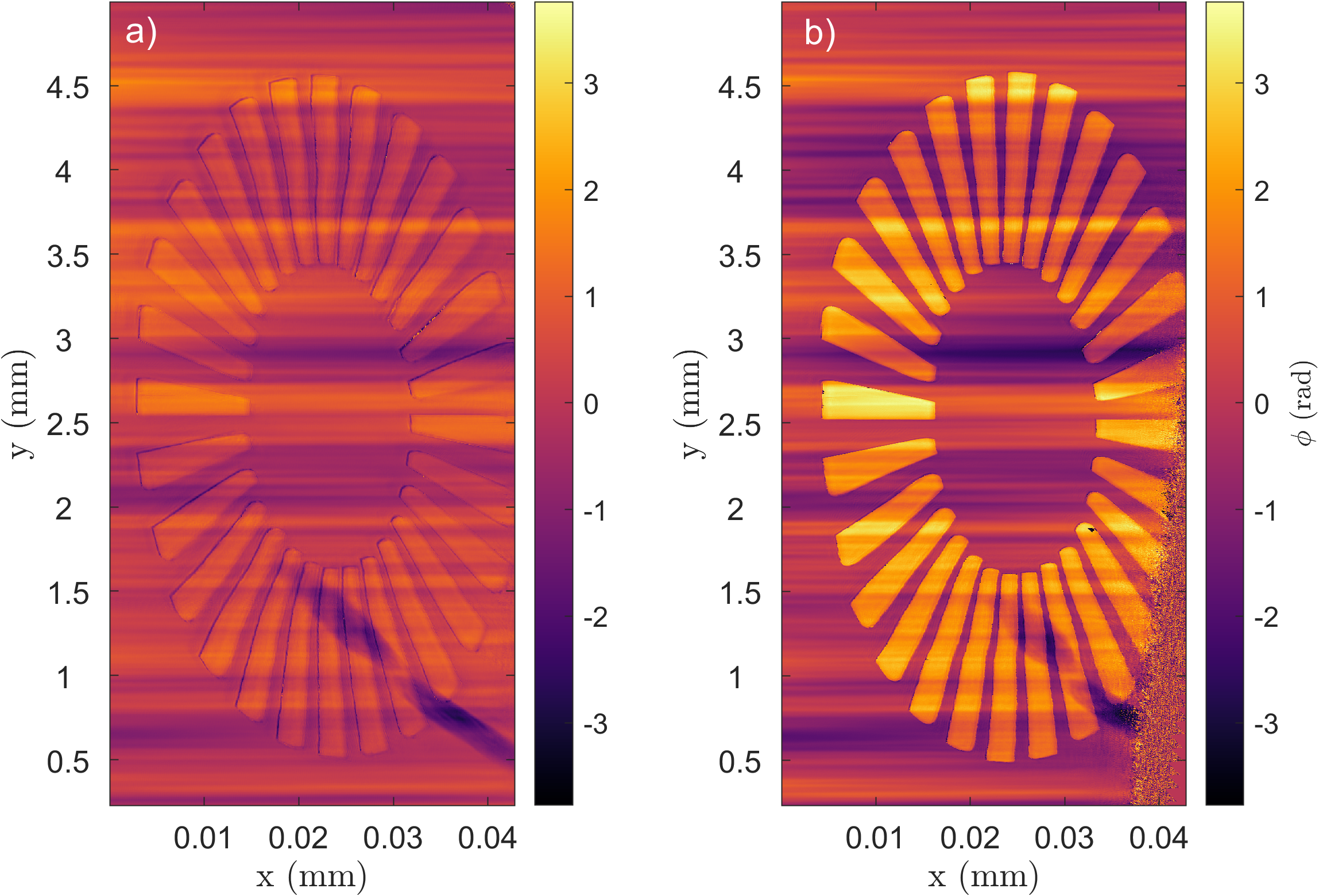}
    \caption{\label{fig:FullSS08deg} Phase of the reconstruction of the full elongated Siemens star test object, taken at a) $\theta=0.6^\circ$, and b) $\theta=0.8^\circ$. Phase given in radians.} 
\end{figure}


 
Fig.~\ref{fig:FullSS08deg} shows the phase contrast images of the fully reconstructed elongated Siemens star structure taken at $\theta=0.6^\circ$ and $\theta=0.8^\circ$, below and above the $\theta_c$ for Au at the experimental X-ray energy of 6.2 keV. The horizontal stripes visible in phase contrast are consistent between varying incidence angle reconstructions and are interpreted as variations in the real surface height of the sample substrate. The stripes appear only in the vertical direction due to the aspect ratio of the images, resulting in a perceived strong 1-D variation in surface height along the y axis parallel to the beam. In reality the surface height variation is uncorrelated in the $x$-$y$ plane. Further, a dark stain-like feature in the bottom half of the star is visible in both images, which is also a real feature of the sample. The feature appears larger in the $\theta=0.6^\circ$ image and we believe this is due to a deposit of a low atomic number material which has a lower refractive index than the sample, and hence appears more clearly at lower incidence angles. The bottom right corner of Fig.~\ref{fig:FullSS08deg} b) shows a small area of noise, corresponding to the edge of the right side of the structure that extends beyond the field of view. 


The relationship between measured phase shift $\phi$ and physical height $h$ of the structure is determined by \mbox{$\phi=\frac{4\pi h}{\lambda} \sin(\theta)$}. Because $\phi$ is computed as the phase of a complex-valued function, it is subject to phase wrapping and its values are restricted within the range of $\pm \pi$, if the height causes a phase shift greater than this, the phase value will wrap around. This implies that an unknown $2\pi n_p$ may be introduced which gives \mbox{$\phi -2\pi n_p=\frac{4\pi h}{\lambda} \sin(\theta)$}. Analagous to the principle of multiple-wavelength interferometry where several wavelengths can be used to increase the range of non-ambiguity in precise length measurements calculated from phase shifts \cite{Dandliker98}, we can use multiple incidence angles to reduce the ambiguity of height measurements. One can disambiguate the phase wrapping by computing heights for a range of $n_p$ values (in this case $1< n_p < 10$ was used). One can then look for the set of $n_p$ values that result in the smallest height difference.


%
In this case, the solution was found to be $n_p=2$ for $\theta=0.6^\circ$ and $n_p=3$ for $\theta=0.8^\circ$, with (unwrapped) phase shift values of 12.5 rad and 16.9 rad, resulting in values of $h=18.9$~nm and $h=19.3$~nm respectively. The recovered structure height is thus 19.1~nm. The contribution of low-frequency errors in the reconstructions can lead to inaccurate estimations of height. In order to avoid this, the reconstructions are filtered through band-passing of the image in the Fourier domain. Further details of this are given in the methods section. The residual error of $h$ as defined by the difference between the observed phase and the phase shift calculated from $\phi -2\pi n_p=\frac{4\pi h}{\lambda} \sin(\theta)$ is found to be 0.4~nm, whereas the angle-dependent uncertainty of height measurements is computed from the standard deviation of the phase measurements of the structure after Fourier domain band-passing and found to be 0.07~nm for $\theta_c = 0.6^\circ$ and 0.06~nm for $\theta_c = 0.8^\circ$. The calculated RMS surface roughness of the substrate is 0.7~nm for $\theta=0.6^\circ$ and 0.5~nm for $\theta=0.8^\circ$ . Further details are discussed in the methods section. This measured height is in excellent agreement with heights of the sample measured by atomic force microscopy which were found to be 19$\pm 3$~nm (3~nm being the mean surface roughness as measured by AFM), showing the excellent sensitivity of phase contrast to nanoscale variations in grazing incidence geometry. 

\subsection*{Simulation}

\begin{figure}[h]
    \includegraphics[width=\columnwidth]{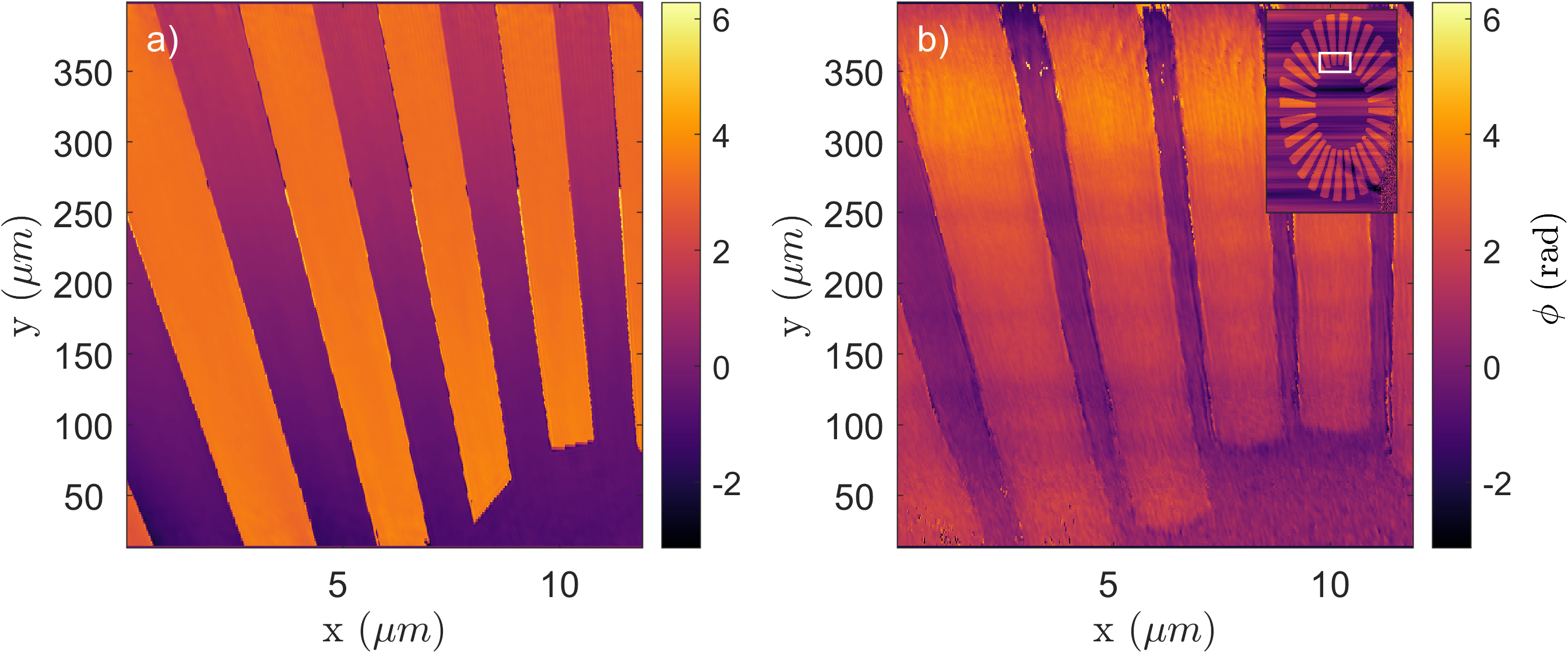}
    \caption{\label{fig:MSvsReal_withinset} a) Multislice simulated phase reconstruction of a smaller area of the test object at $\theta=0.8^\circ$. b) experimental data phase reconstruction of the same area at $\theta=0.8^\circ$. Inset: the ROI on the larger image of the test object.} 
\end{figure}

Fig.~\ref{fig:MSvsReal_withinset} shows the ptychographic reconstruction from diffraction patterns simulated via the multislice propagation method \cite{Li2017,Munro2019} alongside the reconstruction from experimentally collected data for the same area of the Siemens star. Diffraction patterns were simulated via multislice propagation and reconstructed with the same algorithms and parameters as their experimental data counterparts (i.e. multislice simulation is used for the forward model to generate simulated diffraction patterns, but a conventional ptychographic reconstruction with a simple multiplication between object and probe is used in every case). The multislice approach is used as a forward model as it has been shown to be capable of producing arbitrary complex reflections, including evanescent waves\cite{Li2017}. The simulation and the experiment in Fig.~\ref{fig:MSvsReal_withinset} were done at an incidence angle of $\theta=0.8^\circ$. It can be seen that the real structure has rounder edges and less spacing between spokes. These discrepancies are due to the sample fabrication process where limits on edge sharpness are necessary. The observed smaller distance between the spokes in the experimental reconstruction is also a real property of the fabricated sample and not due to experimental accuracy. As expected, the substrate shows little to no phase contrast in the simulated data, whereas there is added variation and horizontal banding across the real reconstruction, which is real surface roughness. Height estimation on multislice simulated data was found to be $n_p=4$ for both $\theta=0.8^\circ$ and $\theta=0.9^\circ$, with (unwrapped) phase shift values of 23.4 rad and 25.8 rad, resulting in values of $h=26.5$~nm and $h=26.1$~nm respectively. The recovered structure height found to be is 26.5~nm, resulting in an overestimation of approximately $1.5$~nm from the simulated height of $25$~nm. We cannot currently explain the cause of this overestimation. Simulated diffraction patterns created by the multislice forward model are in good agreement with experimental data, and as such suggest that multislice simulation represents an adequate forward model for describing the wave-sample interaction in grazing incidence and producing diffraction data for qualitative comparison with experimental data and to validate the reconstructions using a simpler model. Further details of the multislice simulations are in the discussion section.




\subsection*{Resolution Estimation}
Fourier ring correlation (FRC) \cite{van1982arthropod,VanHeel2005,Holler2014} has become a standard method for providing reliable and quantitative estimates of the image resolution across a large number of imaging techniques. To estimate highly anisotropic resolution of images reconstructed from grazing-incidence X-ray ptychographic data, the calculation of FRC has to be decoupled for transverse and longitudinal directions. As such, the FRC is calculated separately for each dimension in the real space image, and the FRC is computed from two separate 1-Dimensional Fourier transforms of each image. 

\begin{figure}[h]
    \includegraphics[width=\columnwidth]{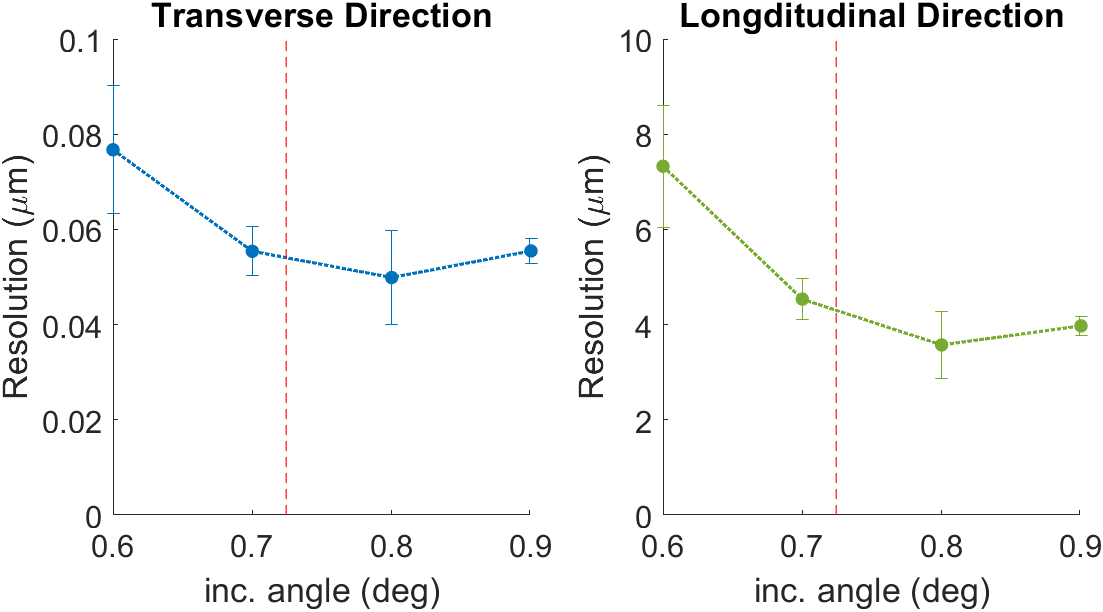}
    \caption{\label{fig:FRC} FRC estimates for each incidence angle observed. The red line indicates the $\theta_{c}$ for Au at the experimental energy. Longditudinal refers to the direction parallel to the beam, whereas transverse is normal to the beam in the $xy$ plane of the sample.} 
\end{figure}

The estimated resolution from FRC is shown in Fig.~\ref{fig:FRC} as a function of incidence angle. The resolution in directions parallel and transverse to the beam approach a minimum near $\theta_{c}$. As expected, the resolution is significantly poorer in the longditudinal direction, and the resolution in both directions becomes poorer as the incidence angle is shifted further away from $\theta_{c}$. While the resolution in the transverse and longditudinal directions is worse for $\theta=0.6^\circ$ than other incidence angles, the FRC resolutions at angle $\theta=0.7^\circ$ and above are within the standard deviation of FRC measurements calculated from several sub-regions of the full image at each incidence angle.

\section*{Discussion}

Measured height differences in reflection are determined by observed phase shifts given by the geometric relation \mbox{$h=\phi/(2k \sin(\theta))$} (where $k = 2\pi/\lambda$). This relationship between $\phi$ and $h$ is based on two assumptions. Firstly, $\phi$ is only determined by topological variation of the sample, as the refractive index, $n$ is assumed to be constant throughout the entire sample. While this is not a requirement of the reconstruction method, the quantitative calculation of height in this work is based on this assumption to simplify the interpretation of phase contrast reconstructions.  Secondly, $\phi$ is considered to be caused by a single scattering upon reflection, without considering any multiple scattering events with higher complexity. 


Whereas these assumptions can be made in the case of transmission of samples where the object transmissivity can be represented by a 2-D function and the sample thickness falls within the depth of focus \cite{Tsai2016,Maiden2012}, in the case of grazing incidence, scattering events with higher complexity cannot generally be ignored. Given the test structures are entirely Au and the Ti and Si substrate layers are well below the 50~nm Au substrate, these layers are assumed to have negligible contribution to the measured signal, as the attenuation length of Au at the experimental energy ranges from approximately 1 to 10~nm over the range of $\theta$ investigated, these are well below the 50nm Au substrate thickness. Therefore a constant refractive index $n$ everywhere is a reasonable assumption in this work. However, this would not be the case for samples that have both chemical as well as topological inhomogeneities, or samples where transmissivity through the structure in grazing incidence cannot be neglected (i.e. samples with significantly lower $\beta$ than Au).  

In particular, for the samples imaged in this work, artefacts appear in phase contrast at the edges of the structure where the phase shift is much larger than $\lambda /2$ per pixel. This causes determination of the true structure height to be less straightforward. Nonetheless the reconstructions in this work are based on this thin-object approximation without accounting for changes to the illumination function throughout the sample and height estimation from phase shift measurements are still found to be in good agreement with AFM measurements. This will likely only hold for relatively simple topologies of non-transmissive material as studied here. The contribution of more complex scattering phenomena as described by higher order Distorted-wave Born Approximation (DWBA) \cite{Sinha1988} terms to the measured phase shift needs to be explored further to better quantify topological contrast in more complex samples, especially for partially transmissive materials.

\section*{Conclusion}

We have demonstrated a grazing-incidence X-ray scattering ptychography experiment over a large area on the millimetre scale in the sample plane with a relatively small number of scan points. The technique is capable of providing nanometre topological resolution through phase contrast. The experiment can be implemented at existing beamlines with existing phase-retrieval algorithms. The applicability of this imaging with excellent surface sensitivity shows promising potential for application of grazing incidence X-ray ptychography for robust large-scale characterization of surfaces and thin films where nanoscale height precision is required. The discrepancy between height estimation from phase contrast measurements and AFM is on the order of 1 nm, whereas the same method using images produced by multislice simulations show good agreement with both experimentally obtained data and with the nominal height used in the input simulation settings. 

Having been in good agreement with experimentally obtained results, the multislice simulations aid in the qualitative interpretation and verification of the experimental data. We have demonstrated that multislice simulations provide a useful forward model for producing diffraction data that can be reconstructed with existing phase-retrieval algorithms. The model of height variation being calculated from phase shift arising from geometric path length difference holds for the structures imaged in this work given they are both chemically homogeneous and a highly absorbing material, however future work is required to more accurately develop a model and reconstruction that takes into account higher-order DWBA terms. 


\section*{Methods}

\begin{figure}[h]
    \includegraphics[width=0.7\columnwidth]{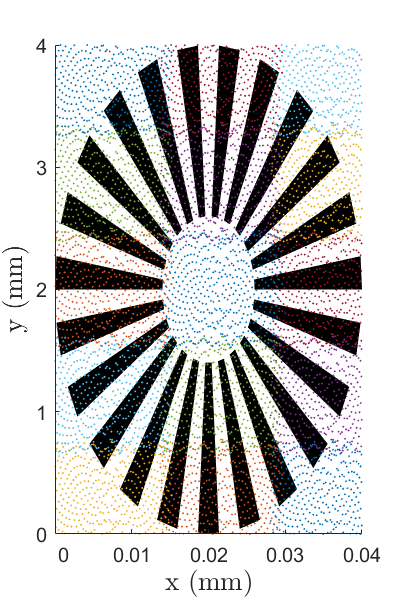}
    \caption{\label{fig:stitchscan} The scanning pattern arrangement with relation to the overall structure to be imaged. Each colour block corresponds to each fermat spiral scan, where each fermat scan contains several hundred points.} 
\end{figure}

\noindent\textbf{Experiment}
The experiment was performed at the cSAXS beamline of the Swiss Light Source (SLS) at the Paul Scherrer Institute in Villigen, Switzerland, using a photon energy of 6.2~keV. The coherent illumination on the sample was defined by a Fresnel zone plate (FZP) made of Au, fabricated by the X-ray nano-optics group at the Paul Scherrer Institute \cite{Gorelick2011}. The FZP had a diameter of 220~$\mu$m and 4~nm outer-most zone width, resulting in a focal length of 99~mm and a focal depth of $\pm80~\mu$m. The reflection geometry of the experiment requires the scanning to be performed in a plane parallel to the sample surface (Fig.~\ref{fig:setup}). In this way, the illumination probe profile on the sample and source-sample distance can be considered constant and ptychographic phase-retrieval can be performed. When referring to the probe illumination size, we usually consider its transverse extent in the sample plane. However, it should be noted that for the incident beam probing a surface at an angle $\theta$, the beam footprint is elongated in the longitudinal direction by a factor of $1/\sin(\theta)$. To keep the probe overlap consistent in both directions, the scanning step size along the grazing axis has to be scaled accordingly. A compact SmarAct hexapod-like positioning system was mounted on top of the scanning piezo-stage to align the sample surface parallel to the scanning plane during acquisition. This was achieved using an interferometric position measurement of the sample height (z-direction). The surface of the sample was used as reflective surface. To align this surface to the scanning plane, the sample stage was continuously moved in a sinusoidal pattern in the $x$-$y$ plane and the measured displacement in the $z$-direction was minimized by adjusting the sample tilt using the SmarAct positioning system.. The height variation is minimized independently for both $x$ and $y$ through fine adjustments of the smaract stage. Scans were performed in an elongated fermat spiral pattern \cite{Huang2014}, where the step size of each scan were elongated by a factor of $1/\sin(\theta$) in the direction parallel to the beam propagation. Acquisition times were 0.2 s per position. Using a 4~$\mu$m size beam in the transverse direction, an effective area of 40 $\times$ 500 $\mu \textrm{m}^2$ was covered per fermat spiral. A Pilatus 2M detector with a pixel size of 172 $\mu$m was used for collection of diffraction data at a sample to detector distance of 7.36 m. Due to the limited range of the piezo system, the scan was split into several subscans as shown in Fig.~\ref{fig:stitchscan}, where each fermat scan was performed with the piezo stage after coarser translations with the hexapod. At the X-ray photon energy of 6.2 keV, $\theta_{c}$ for Au is $\theta_c \approx 0.72^\circ$. Samples were imaged at 0.6, 0.7, 0.8, and 0.9$^\circ$, a range of angles below and above $\theta_{c}$.


\noindent\textbf{Sample fabrication}
The patterns were created on a $\langle 001 \rangle$ Si wafer. Initially, a $10$~nm layer of Ti is evaporated onto the wafer, followed by a $50$~nm Au layer, using e-beam evaporation in a Temescal FC-2000 tool with a deposition rate of 2 Å/s. The Ti layer is necessary to ensure good adhesion of Au on the substrate. After the initial Ti/Au bilayer, the wafers were spun with $1.5~\mu$m positive photoresist AZ-MIR701 and the structures were patterned using UV lithography and further developed in a TMAH solution. The wafers are then placed in the e-beam evaporated and an additional layer of $20$~nm of Au was evaporated to create the final patterns. The thickness of the Au deposition is controlled using a quartz crystal monitoring system. The resist was then lifted-off using a solvent solution (Remover 1165) at room temperature in an ultrasonic bath for approximately 5 minutes, leaving behind the patterned structures. The wafer was then rinsed with isopropanol for a further 5 minutes, followed by DI water, and then finally air dried.

\noindent\textbf{Numerical simulations}
Wave interaction with matter in grazing-incidence geometry cannot be approximated with the first Born approximation due to complex scattering phenomena and the more general DWBA is usually considered \cite{Sinha1988}. Scattering amplitude in DWBA is calculated as a coherent sum of scattering amplitudes contributing from different mixtures of refraction and scattering events. In general, four main scattering events (referred to as channels) with highest contributions are taken into account \cite{Vineyard82,Sinha1988}. In such case, relating the scattering amplitude to electron density of the specimen is usually done by fitting a theoretical model to the experimental data taken in reflectivity by considering the contribution of these four scattering events. 
Recently, it was shown that for the wave incident on the specimen under grazing angle, the so-called multislice propagation can model a range of complex scattering phenomena, such as standing and evanescent waves in the vicinity of the probed surface \cite{Li2017,Munro2019}. In this approximation, a specimen described in terms of complex refractive indices is divided into a set of thin slices. The transmission of the wave through each slice satisfies the projection approximation and propagation between slices is modeled using the angular-spectrum non-paraxial propagator \mbox{$\psi_{j+1}=\mathcal{F}^{-1}\{\mathcal{F}\{\psi_j\}\cdot \exp[-i\frac{2\pi\Delta z}{\lambda}\sqrt{1-\lambda^2(u_x^2+u_y^2)} ] \}$} where $\mathcal{F}$ and $\mathcal{F}^{^-1}$ are the Fourier transform and its inverse, $\psi_{j}$ is the $j_{th}$ wave in the simulation, $\Delta z$ is the thickness of a slice, $\lambda$ is the wavelength, and $u_{xy}$ are spatial frequencies. Wave-propagation with multislice approximation has been widely used as a forward model for simulating diffraction data in various applications \cite{Li2017}. 

Using the method outlined in  \citet{Li2017}, the goal of the multislice simulation was to replicate the real experiment faithfully. All layers of the substrate material were simulated using the tabulated complex refractive index of each material. The virtual sample was constructed by importing the same pattern definition file used to create the physical sample, and discretizing it onto the simulation grid. The motor positions from the real experiment were imported and translated into movements of the virtual sample. Finally, the sample phantom was reconstructed from the simulated diffraction patterns using the same pipeline as used for the data from the physical experiment, with the same parameters. Reconstructions were made without adding noise to the diffraction patterns to attempt to simulate the experiment under ideal conditions.

The simulation volume is discretized into a 5~nm voxel grid in the directions transverse to the propagation direction, and 200~nm along the propagation direction. This is chosen because the interaction length in grazing incidence is much longer along the propagation direction, and in real reconstructions, resolution along the propagation direction is decreased. As a result, the resolution may be relaxed in this dimension for the simulation. The input wave chosen for the simulations is a reconstructed probe from experimental data from the cSAXS beamline, which is the wave field at the plane where it interacts with the sample, downstream of the focal point. This allows for ptychographic reconstructions of simulated data to use the reconstructed probe from previous scans as an initial guess, which greatly helps convergence. 

The incoming wave is tilted to match grazing incidence geometry before propagating orthogonally through the simulated volume, each slice being in the plane normal to the direction of propagation, and then tilted again after interaction with the sample so the final exit wave is once again orthogonal to the detector. The simulation size is on the order of 1 to 4 $\times10^3$ slices, with each slice being approximately $1700 \times 800$ voxels.

\noindent\textbf{Reconstruction}
Ptychographic reconstructions were completed using Ptychoshelves \cite{Wakonig:zy5001}. A square area of 182 pixels around the center of the reflected beam from each far-field scattering pattern was used as the input for reconstructions. For reconstructions covering a large field of view, several overlapping Fermat spirals are combined together into a single ptychographic reconstruction with a shared object \cite{Sicairos2014,Huang2014}. 

Two methods are used in succession for solving the ptychographic reconstruction, firstly the difference map algorithm \cite{Thibault2008}, followed by least squares maximum likelihood method using compact sets (LSQ-MLc) \cite{Odstrcil18}. As a stopping criterion, the number of iterations for each method is fixed at 300. Collected scattering patterns, which, in the plane of the detector are tilted in Fourier space, undergo a coordinate transform to have uniform spacing in Fourier space prior to solving. This process is more commonly referred to as tilted plane correction \cite{Gardner:12}. After reconstruction, the samples are corrected for a linear phase-ramp by fitting a 2-D plane to bare regions of the substrate across the image \cite{Guizar-Sicairos:11}, and 2-D phase unwrapping is performed.

\noindent\textbf{Resolution estimation}
1-D decoupled FRC involves performing separate 1-D Fourier transforms along the transverse and parallel directions of the beam propagation respectively. This allows for a separation of spatial frequencies between the lower frequency range of the longditudinal axis, and the higher frequency range of the transverse axis. A pair of images of the same area were taken for each incidence angle, and the correlation of their Fourier transforms for a single axis at a given incidence angle are determined. For FRC calculations, a finer structure, on the same wafer, fabricated and made in the same method as all other structures, but with much smaller features than the Siemens star in the x-y plane, was used for FRC calculations. All images used for FRC have tilted-plane correction applied to their ptychographic reconstructions prior to FRC. 

Height measurements, which are encoded in phase shift measurements in these images, are dependent on spatial frequency and therefore biased by low frequency noise in the reconstructions. Band-passing is therefore required to achieve an accurate estimate of height. To achieve this, phase images are band-passed using a top-hat filter in the frequency domain to within a range of one decade of signal, chosen to be between the 1 $\times$ FRC resolution and 10 $\times$ FRC resolution estimated for each incidence angle. The standard deviation of phase measurements from two independent sets of data within the ROI after band-passing then provides an estimate of the repeatable precision of phase measurements and height sensitivity for each incidence angle. The standard deviation between these two sets of measurements provide an estimate of the achievable precision of height measurements along the z-axis, whereas FRC estimates provide the error along $x$ and $y$, i.e. the resolution along transverse and longitudinal directions in the plane of the sample. 




\begin{acknowledgments}
This study was partially funded from the European Union’s Horizon 2020 research and innovation programme under the Marie Skłodowska-Curie grant agreement No. 765604 (MUMMERING). We also wish to acknowledge support from the Villum Experiment Programme and the Velux Foundations. We gratefully acknowledge the contribution of Professor Ole Hansen from DTU Nanolab to sample design and supervision of manufacturing.
\end{acknowledgments}

\section*{Competing interests}
The authors declare that they have no conflict of interest

\bibliography{giptycho}

\end{document}